\documentclass{kapproc} 

\usepackage{procps}

\usepackage[dvips]{graphicx}

\upperandlowercase

\normallatexbib 

\def \be {\begin{equation}}
\def \ee {\end{equation}}
\def \beq {\begin{eqnarray}}
\def \eeq {\end{eqnarray}}
\def \bean {\begin{eqnarray*}}
\def \eean {\end{eqnarray*}}

\def \part {{\partial}}

\begin{document}





\articletitle[Coupling of radial and non-radial oscillations of relativistic stars]
{Coupling of radial and non-radial oscillations of Neutron stars.}

\author{Andrea Passamonti,\altaffilmark{1} Marco Bruni,\altaffilmark{1}
Leonardo Gualtieri,\altaffilmark{2} and Carlos F. Sopuerta\altaffilmark{1,3}}

\altaffiltext{1}{ICG, University of Portsmouth, Portsmouth PO1 2EG, Britain}

\altaffiltext{2}{Dipartimento di Fisica ``G. Marconi'', Universit\`a di Roma
``La Sapienza'' and Sezione INFN ROMA 1, piazzale Aldo Moro 2, I--00185 Roma,
Italy}

\altaffiltext{3}{IGPG and CGWP, The Pennsylvania State University,
University Park, PA 16802, USA}


\begin{abstract}
This is a progress report on our study of the coupling of first-order radial and
non-radial relativistic perturbations of a static spherical star.
Our goal is to investigate the effects of this coupling on the gravitational
wave signal of neutron stars.  In particular, we are looking for the existence
of resonances and parametric amplifications, changes in the damping
time of non-radial oscillations, etc.   To that end, we have developed
a formalism that introduces gauge invariant quantities to describe
the coupling.  Their equations have the same structure as
the equations for first-order non-radial perturbations plus some
source terms, which makes them very appealing for time domain studies.
\end{abstract}


\section*{Introduction}
Neutron stars are important gravitational wave sources both as
isolated objects or in binary systems. The analysis of their
gravitational radiation could open up a direct window on their
interior, possibly revealing details on the equation of state of
nuclear matter, the dynamics of the crust-mantle interaction and
the inner superfluid/superconducting core. An accurate modelling
of sources is crucial to the final end of providing templates of
the wave form and spectrum of the signal for the analysis of data
that comes from the new generation of gravitational wave
detectors.

Linear perturbations and instabilities of neutron stars have been
studied for a long time~\cite{Andersson:2002ch} but
relatively little is known of non-linear dynamical effects (see
\cite{Passamonti} for references) and therefore second order
studies may help to understand known problems and even reveal a
new phenomenology.

Here we outline our work in~\cite{Passamonti}, where we introduced
a framework to study the coupling of radial and non-radial
first-order relativistic perturbations of static spherical stars.
The goal is to study effects of this coupling (possible resonances,
parametric amplification, etc.) that could make a significant
impact in the gravitational wave signal of neutron stars.


\section{Perturbative Framework} \label{framework}

Using 2-parameter relativistic perturbation theory~\cite{Bruni:2002sm,Sopuerta:2003rg}
we parametrize separately the radial and non-radial perturbations.  This allows
us to split second order perturbations into three differentiated types: second-order
radial and non-radial and the coupling between them.   We are interested
in the third type.

The basic ingredients of our perturbative framework are: (i) {\em The background
model for the star}.  We use a Tolman-Oppenheimer-Volkov model, that is, an equilibrium
perfect-fluid spherically symmetric configuration.  (ii) {\em First-order
radial perturbations} (see~\cite{Kokkotas:2002ng} and references therein).  These
are perturbations preserving the spherical symmetry of the background model.
They can be described by just three functions: two matter variables (in our case
we choose the enthalphy and the fluid velocity radial component perturbations)
and a metric variable.  They obey a system of evolution
equations containing only first-order time derivatives and subject to a
constraint, the Hamiltonian constraint, which
can be used to monitor the accuracy of a time-domain integration.
(iii) {\em First-order non-radial perturbations}.   Due to the symmetry of
the background, these perturbations can be expanded in terms of (tensor)
harmonics, so that the angular dependence is explicitly separated.
Depending on how they behave under parity transformations they are
called {\em polar} (even) or {\em axial} (odd) perturbations.
Every harmonic (the {\em monopole} part would correspond to the radial
perturbations, so it is not consider here)
can be described in terms of gauge-invariant variables by using the formalism
developed in~\cite{Gerlach:1979rw,Gundlach:1999bt,Martin-Garcia:2000ze}.
These perturbations have been extensively studied in the literature due
their interest in relation with the gravitational radiation produced
by star oscillations.  Most of the works in the past were done in
the frequency domain~\cite{Thorne:1967th,Detweiler:1985dl,Chandrasekhar:1991fi}.
However, recent works have studied them by using a time-domain
approach~\cite{Allen:1998xj,Ruoff:2001ux,Nagar:2004ns}.
(iv) {\em Coupling of radial and non-radial perturbations}.
As we have mentioned above, these perturbations are a part of the second-order
perturbations of our background.  The part that is generated by the coupling
of the radial and non-radial first-order perturbations.  This is the
sector of the second-order perturbations that we have to study in order to
look for the physical phenomena we described in the abstract and introduction,
and it is the subject of~\cite{Passamonti}.   The structure of these perturbations
is very particular.  By pure inspection of Einstein's equations one can
see that they are generated by source terms that can be expressed as a sum of
products of radial and non-radial first order 
perturbations.  This means that they
can be also expanded in (tensor) harmonics, which makes the analysis much
simpler than the analysis of the whole set of second-order perturbations.
Moreover, in~\cite{Passamonti} we were able to show that one can also
have a gauge-invariant description just by extending in an appropriate
way (in particular, by fixing the gauge for the radial perturbations)
the formalism for non-radial perturbations introduced
in~~\cite{Gerlach:1979rw,Gundlach:1999bt,Martin-Garcia:2000ze}.

Going into more detail about the structure of the equations for the
coupling perturbations, we have seen in~\cite{Passamonti} that these
perturbations, once decomposed in harmonics, obey the same equations
as non-radial perturbations do, with the only difference that for the
coupling terms we have source terms which, as we mentioned before,
can be written as the sum of products of radial and non-radial
perturbations.   This structure has very important consequences for
practical purposes, in particular for the numerical integration of the
equations.  Indeed, given a numerical code capable of evolving the
non-radial perturbations we can construct a code evolving the coupling
perturbations just by adding the sources.   Having this interesting
property in mind, an appropriate way of formulating (choice of variables
and equations) the equations for non-radial perturbations would be
the one choosen in~\cite{Nagar:2004ns}, where the Hamiltonian constraint
(an elliptic-type equation) is used to solve for one of the perturbative
variables instead of using an evolution equation.  Apart from the
obvious interest that this procedure has (we make sure that constraints
are preserved during the evolution), it has an extra interest thinking
on solving also for the coupling perturbations: If we do not solve for
the Hamiltonian constraint the errors produced by its violation would
increase since we would accumulate the ones coming from the integration
of the non-radial perturbations with the ones after solving for the
coupling.  Therefore, solving for the Hamiltonian constraint can improve
substantially the accuracy of the calculations.  As it has been shown for
the case of non-radial perturbations it can estimate damping times and
mode frequencies with an accuracy comparable to frequency domain calculations 
\cite{Nagar:2004ns}.
Hence, the structure of the system of equations governing the stellar interior 
is given by a \emph{gravitational wave equation} for the
non conformal-flat metric perturbation  $ S $, a
\emph{sound wave equation} for a fluid perturbation $ H $ 
(coincident with the enthalpy perturbation in some particular gauges), 
and finally the \emph{Hamiltonian constraint} mentioned above, used to 
update the value of the conformal-flat metric perturbation 
$k $  at every time-step, 
\beq
- S_{,tt} & + & e^{2 (\Phi - \Lambda)}
S_{,rr} + ...... =  e^{2  \Phi} {\cal S}_{S}\,. \label{GW11} \\
- H_{,tt}   & + & \bar{c}_s^2 e^{2 \left(\Phi -\Lambda \right)}
H_{,rr}   + ...... = e^{2 \Phi}   {\cal S}_{H}\,, \label{SW11}  \\
k_{,rr} & + & \frac{2}{r\bar{c}_s^2} \left(\Lambda_{,r}+\Phi_{,r}\right)
H + ...... = {\cal S}_{Hamil} \,, \label{Ham11}
\eeq
where ${\cal S}_{S}$,${\cal S}_{H}$,${\cal S}_{Hamil}$ are the source terms 
containing the product of first order radial/nonradial perturbations 
\cite{Passamonti}. 

To sum up, our framework to study the coupling of radial and non-radial
oscillations leads to a hierarchy of equations (from the background to
the coupling terms).  In order to solve them in the time domain we have
to pay attention to a number of other important issues.  Of particular relevance
are the boundary conditions:  We need to impose the regularity of the
perturbations at the origin, the vanishing of the Lagrangian perturbation
of the pressure at the star surface, and the continuity of metric perturbations
at the surface (junction conditions), connecting with the exterior, which
can be described by the corresponding Zerilli equation, which will propagate
the gravitational wave signal.  Then, we can use the well-known black-hole
perturbation machinery to compute the energy and angular momentum
that has been radiated away.    A discussion of all these issues can also
be found in our first work~\cite{Passamonti}.

\section{Future work}

We are presently working in the construction of numerical codes 
for the time domain integration of the equations derived in
\cite{Passamonti}.   In this work we focused on polar perturbations,
which are the most relevant for stars, but we are also
exploring the axial case.  On the other hand, this work can also be considered 
as a step towards
a more comprehensive study of second order perturbations of compact stars
and mode coupling.

\begin{acknowledgments}
This work has been partially
supported by the EU (Research Training Network
contract HPRN-CT-2000-00137).
CFS was supported by EPSRC and presently by NSF grants PHY-9800973 and
PHY-0114375.
\end{acknowledgments}








{




\begin{chapthebibliography}{1}

\def\lcitebracket{[} \def\rcitebracket{]}

\bibitem{Andersson:2002ch} N. Andersson, Class. Quant. Grav. {\bf 20}, R105 (2003).

\bibitem{Passamonti} A. Passamonti, M. Bruni, L. Gualtieri, and C.F. Sopuerta,
(2004), gr-qc/0407108.











\bibitem{Bruni:2002sm} M. Bruni, L. Gualtieri, and C.F. Sopuerta, Class. Quant. Grav.
{\bf 20}, 535 (2003).

\bibitem{Sopuerta:2003rg} C.F. Sopuerta,  M. Bruni, and  L. Gualtieri,
Phys. Rev. D {\bf 70}, 064002 (2004).

\bibitem{Kokkotas:2002ng} K.D. Kokkotas and J. Ruoff (2002), gr-qc/0212105.

\bibitem{Gerlach:1979rw} U.H. Gerlach and U.K. Sengupta, Phys. Rev. D {\bf 19},
2268, (1979).


\bibitem{Gundlach:1999bt} C. Gundlach and J.M. Martin-Garcia,
Phys. Rev. D {\bf 61}, 084024 (2000).

\bibitem{Martin-Garcia:2000ze} J.M. Martin-Garcia and C. Gundlach,
Phys. Rev. D {\bf 64}, 024012 (2001).








\bibitem{Thorne:1967th} K.S. Thorne and A. Campolattaro, Astrophys.J. {\bf 149}, 591 (1967).


\bibitem{Detweiler:1985dl} S. Detweiler and L. Lindblom, Astrophys.J. {\bf 292}, 12 (1985).

\bibitem{Chandrasekhar:1991fi} S. Chandrasekhar and V. Ferrari, Proc. Roy. Soc.
(London) A {\bf 432}, 247 (1991).


\bibitem{Allen:1998xj} G. Allen, N. Andersson, K.D. Kokkotas, and B.F. Schutz,
Phys. Rev.D {\bf 58}, 124012 (1998).


\bibitem{Ruoff:2001ux} J. Ruoff, Phys. Rev.D {\bf 63}, 064018 (2001).


\bibitem{Nagar:2004ns} A. Nagar, G. Diaz, J.A. Pons, and J.A. Font,
Phys. Rev. D {\bf 69}, 124028  (2004);
A. Nagar and G. Diaz,
(2004), gr-qc/0408041.




\end{chapthebibliography}
\end{document}